\documentclass[10pt]{iopart}
\usepackage{graphicx}
\usepackage{xspace}				
\usepackage{colortbl}				

\newcommand{\esa}{\textsl{ESA}\xspace}

\newcommand{\lisa}{\textsl{LISA}\xspace}
\newcommand{\lpf}{\textsl{\mbox{LPF}}\xspace}
\newcommand{\ltp}{\textsl{LTP}\xspace}
\newcommand{\nasa}{\textsl{NASA}\xspace}

\begin{document}


\jl{6}

\title[On-ground tests of \textsl{LISA PathFinder}\ldots]{On-ground tests
of \textsl{LISA PathFinder} thermal diagnostics system}

\author{A Lobo$^{1,2}$\footnote[3]{To whom correspondence should be
addressed.}, M Nofrarias$^2$, J Ramos-Castro$^3$ and J Sanju\'an$^2$}

\address{$^1$ Institut de Ci\`encies de l'Espai, {\sl CSIC}}

\address{$^2$ Institut d'Estudis Espacials de Catalunya ({\sl IEEC\/}), Edifici
	{\sl Nexus}, Gran Capit\`a~2--4, 08034 Barcelona, Spain}

\address{$^3$ Departament d'Enginyeria Electr\`onica, {\sl UPC},
	Campus Nord, Edif.\ C4, Jordi Girona 1--3, 08034 Barcelona, Spain
	\ead{lobo@ieec.fcr.es}}

\date{\today}

\begin{abstract}
Thermal conditions in the \ltp, the \lisa Technology Package, are
required to be very stable, and in such environment precision temperature
measurements are also required for various diagnostics objectives.
A sensitive temperature gauging system for the \ltp is being developed
at IEEC, which includes a set of thermistors and associated electronics.
In this paper we discuss the derived requirements applying to the
temperature sensing system, and address the problem of how to create
in the laboratory a thermally quiet environment, suitable to perform
meaningful on-ground tests of the system. The concept is a two layer
spherical body, with a central aluminium core for sensor implantation
surrounded by a layer of polyurethane. We construct the insulator
transfer function, which relates the temperature at the core with
the laboratory ambient temperature, and evaluate the losses caused
by heat leakage through connecting wires. The results of the analysis
indicate that, in spite of the very demanding stability conditions,
a sphere of outer diameter of the order one metre is sufficient.
We provide experimental evidence confirming the model predictions.
\end{abstract}
%
\submitto{\CQG}
%

\section{Introduction
\label{sec.1}}

\lisa Pathfinder (\lpf) is an \esa mission, with \nasa contributions,
whose main objective is to put to test critical parts of \lisa (Laser
Interferometer Space Antenna), the first space borne gravitational wave
(GW) observatory~\cite{bender}. The science module on board \lpf is the
\lisa Technology Package (\ltp)~\cite{lpfall}, which basically consists
in two test masses in nominally perfect geodesic motion (free fall), and
a laser metrology system; this one detects \emph{residual deviations} of
the test masses' actual motion from the ideal free fall, to a given level
of accuracy~\cite{gerhar}.

In order to ensure that the test masses are not deviated from their
geodesic trajectories by external (non-gravitational) agents, a so
called Gravitational Reference System (GRS) is used~\cite{rita}. This
consists in position sensors for the masses which send signals to a set
of micro-thrusters; the latter take care of correcting as necessary the
spacecraft trajectory, so that at least one of the test masses remains
centred relative to the spacecraft at all times. The combination of the
GRS plus the actuators is known as \emph{drag-free} subsystem\footnote{
The term \emph{drag-free} dates back to the early days of space
navigation, when it was used to name a trajectory correction system
designed to compensate for the effect of atmospheric drag on satellites
in low altitude orbits.}.

The \emph{drag-free} is of course a central component of \lisa, and
needs to be operated at extremely demanding levels of accuracy. The
laser metrology system should then be sufficiently precise to measure
relative test mass deviations. The overall level of noise acceptable
for \lisa is defined in terms of rms acceleration spectral density,
and has been set to~\cite{bender}
\begin{equation}
 S_{a,{\rm LISA}}^{1/2}(\omega)\leq 3\!\times\!10^{-15}\,\left[
 1 + \left(\frac{\omega/2\pi}{3\ {\rm mHz}}\right)^{\!\!2}\right]\,
 {\rm m}\,{\rm s}^{-2}/\sqrt{\rm Hz}
 \label{eq.1}
\end{equation}
in the frequency range
$0.1\,{\rm mHz}\leq\omega/2\pi\leq 100\,{\rm mHz}$.
This is equivalent to $S_h^{1/2}$\,$\sim$\,$4\times 10^{-21}$
Hz$^{-1/2}$, with the same frequency dependence.

\lpf is conceived, as mentioned above, as an in-flight test of key
technologies for \lisa. Top level performance requirements for \lpf
have however been relaxed by an order of magnitude relative to \lisa,
both in noise amplitude and in frequency band, to still challenging
goals~\cite{toplev}:
\begin{equation}
 S_{a,{\rm LPF}}^{1/2}(\omega)\leq 3\!\times\!10^{-14}\,\left[
 1 + \left(\frac{\omega/2\pi}{3\ {\rm mHz}}\right)^{\!\!2}\right]\,
 {\rm m}\,{\rm s}^{-2}/\sqrt{\rm Hz}
 \label{eq.2}
\end{equation}
in the frequency range $1\,{\rm mHz}\leq\omega/2\pi\leq 30\,{\rm mHz}$.
We shall be referring to this frequency band as the \ltp Measuring
Bandwidth (MBW) in the sequel.

Equation~\eref{eq.2} gives the \emph{global} noise budget. This is made up
of contributions from various perturbative agents, both of instrumental
and environmental origin. One of these is \emph{temperature fluctuations},
for which a stability requirement has been set to
\begin{equation}
 S_{T}^{1/2}(\omega)\leq 10^{-4}\,{\rm K}/\sqrt{\rm Hz}\ ,
 \quad 1\,{\rm mHz}\leq \omega/2\pi \leq 30\,{\rm mHz}
 \label{eq.3}
\end{equation}
in order to comply with~\eref{eq.2} under suitable noise apportioning
criteria ---see section~\ref{sec.11}.

Because temperature stability is important, a decision has been taken to
place high precision thermometers in several strategic spots across the
\ltp ---as part of what is called \emph{Diagnostics Subsystem}~\cite{lobo}
\footnote{
The Diagnostics Subsystem of the \ltp also includes magnetometric
measurements and a charged particle flux detector.}. Such high precision
temperature measurements will be useful to identify the fraction of the
total system noise which is due to thermal fluctuations only, and this
will in turn provide important debugging information to assess the
performance of the \ltp~\cite{nored}.

The best temperature sensors for our purposes are electric devices
whose ohmic resistance varies with temperature. We have chosen to use
\emph{thermistors}, which are semiconductor resistors whose resistance
decreases as their temperature increases~\cite{ntcs}, because they have
a rather steep sensitivity curve, and this suits our needs. Such sensors
need additional electronic circuitry to bias them and acquire data. The
entire chain of sensors plus electronics must of course be tested in
ground before boarding, and this requires suitably stable environmental
conditions in the first place.

This paper addresses the problem of which are these conditions, and how
to implement them for a reliable laboratory test of the \ltp temperature
sensors and electronics. It is organised as follows: in section~\ref{sec.11}
we review and quantify the identified sources of thermal noise which
result in the budget given in equation~\eref{eq.3}. In section~\ref{sec.12}
we discuss how the stability requirement results in a requirement on the
temperature system performance, and extend the argument to make precise
the environmental conditions which must be met in the laboratory test.
In section~\ref{sec.2} we present the physical hypotheses, lay down the
mathematical model of the proposed insulation scheme, and find an analytic
solution to the equations. We then address in section~\ref{sec.3} the
numerical implications of the above for realistic situations, including
a discussion of heat leakage through connecting wires. Section~\ref{sec.4}
deals with the experimental verification of the model predictions, and
section~\ref{sec.5} summarises our conclusions and future prospects.
Some supplementary mathematical detail is provided in an appendix.

\section{Thermal disturbances in the \textsl{LTP}
\label{sec.11}}

Temperature fluctuations inside the \ltp result in noisy readout at
the interferometer output port ---the \emph{phasemeter}. The reason
for this is that there are components in the \ltp whose behaviour is
sensitive to temperature changes, as we shall describe in this section.

As a rule of thumb, the total contribution of thermal noise to the
total acceleration noise, equation~\eref{eq.2}, should not exceed
10\,\%. We thus require that
\begin{equation}
 S_{a,{\rm thermal}}^{1/2}(\omega)\leq 3\!\times\!10^{-15}\,\left[
 1 + \left(\frac{\omega/2\pi}{3\ {\rm mHz}}\right)^{\!\!2}\right]\,
 {\rm m}\,{\rm s}^{-2}/\sqrt{\rm Hz}
 \label{eq.3a}
\end{equation}
for frequencies within the \ltp MBW. This assumption is in fact somewhat
conservative, as the Project Engineers have estimated that more than
twice this value is actually compliant with the overall \ltp noise
budget~\cite{ASU2007}. We shall however adopt equation~\eref{eq.3a}
as reference to ensure we are playing on the safe side.

For the sake of clarity, we consider separately the influence of
temperature on the \textsl{GRS} and on the Optical Metrology System
(\textsl{OMS}).

\subsection{Noise effects inside the \textsl{GRS}
\label{sec.11-1}}

Temperature differences between the walls of the electrode housing cause
differential pressures on opposite faces of the test masses, which in
turn result in net forces on them, hence in noise at the phasemeter.

Three different mechanisms have been identified whereby temperature
fluctuations distort the \ltp readout: radiation pressure, radiometer
effect and outgassing. Let us briefly describe each of them, and
quantitatively estimate their respective contributions to thermal noise.

\subsubsection{{\sf Radiation pressure}
\label{sec.11-11}}

\begin{figure}[t]
\centering
\includegraphics[width=7cm]{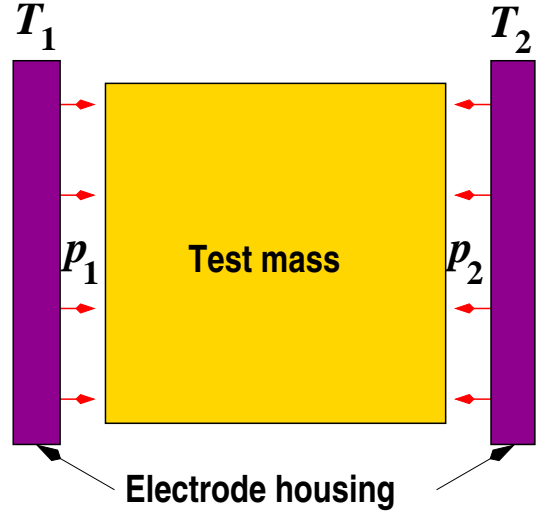}
\caption{Schematics of the effect of different pressures on opoosite faces
of a test mass. The physical origin of pressure differences is explained
in the text. \label{fig.0}}
\end{figure}

A body at any (absolute) temperature $T\/$ emits thermal radiation.
This exerts pressure on any surfaces the radiation hits. According
to standard electromagnetic theory, such pressure is given by
\begin{equation}
 p_{\rm e.m.} = \frac{4}{3}\,\frac{\sigma}{c}\,T^4
 \label{eq.3b}
\end{equation}
where $\sigma$\,=\,5.67$\times$10$^{-8}$\,Wm$^{-2}$K$^{-4}$ is the
Stefan-Boltzmann constant, and $c\/$ is the speed of light. Thus, if
there are temperature fluctuations around the test mass, a noisy net
force will appear on it ---see Figure~\ref{fig.0} for a graphical
display. The effect can be easily quantified making use of
equation~\eref{eq.3b}:
\begin{equation}
 \Delta p_{\rm e.m.} = \frac{16\,\sigma}{3\,c}\,T^3\Delta T
 \label{eq.3c}
\end{equation}
where $\Delta p\/$ and $\Delta T\/$ make reference to differences of
pressure and temperature between the sides of the test mass. Associated
acceleration noise is hence obtained multiplying the above by the test
mass surface area, $\ell_{\rm TM}^2$ and dividing by its mass $m_{\rm TM}$:
\begin{equation}
 \Delta a_{\rm e.m.} =
 \frac{16\,\ell_{\rm TM}^2\sigma}{3\,m_{\rm TM}c}\,T^3\Delta T
 \label{eq.3d}
\end{equation}

\subsubsection{{\sf Radiometer effect}
\label{sec.11-12}}

This is an effect which happens in rarefied gas atmospheres, its name
historically coming from its association with the theory of Crookes's
lightmill radiometer. In low pressure atmospheres, where the gas particles
have a mean free path well in excess of the dimensions of the containing
vessel, equilibrium conditions do not happen when pressure is uniform,
but rather when the ratios of pressure to square root of temperature
equal one another. Or~\cite{kikro},
\begin{equation}
 \frac{p_1}{\sqrt{T_1}} =
 \frac{p_2}{\sqrt{T_2}}
 \label{eq.3e}
\end{equation}

The pressure gradient is readily obtained from this expression, and
thence the associated test mass acceleration:
\begin{equation}
 \Delta a_{\rm radiometer} = \frac{1}{2}\,
 \frac{p\,\ell_{\rm TM}^2}{m_{\rm TM}}\,
 \frac{\Delta T}{T}
 \label{eq.3f}
\end{equation}

\subsubsection{{\sf Outgassing}
\label{sec.11-13}}

Outgassing is one of the causes of the presence of gas within the walls
of the {\sl GRS}. In the present contetx, outgassing problems actually
derive from \emph{temporal fluctuations} in its rate, which once more
result in pressure fluctuations, thence in noise. Outgassing rates are
very strongly dependent on materials and geometry, and the theoretical
analysis of the problem is by no means a simple one. The issue has also
attracted attention of people from other space missions~\cite{nobili},
and further experimental work appears to be necessary to reliably
assess the impact of this phenomenon. Partial evidence has however
been gathered that outgassing might be in practice a small effect in
the \ltp~\cite{ludo}. We shall therefore omit any further consideration
of this phenomenon here.

\subsubsection{{\sf Total thermal noise in the GRS}
\label{sec.11-14}}

If we make the assumption that radiometer and radiation pressure
fluctuations are uncorrelated, and neglect outgassing, then
equations~\eref{eq.3d} and~\eref{eq.3f} are added quadratically,
and hence the spectral densities of acceleration and temperature in
the {\sl GRS\/} are related by
\begin{equation}
 \hspace*{-1 cm}
 S_{a,{\rm thermal\ GRS}}^{1/2}(\omega) = \sqrt{
 \left(\frac{16\,\ell_{\rm TM}^2\sigma}{3\,m_{\rm TM}c}\,T^3\right)^{\!\!2} +
 \left(\frac{p\,\ell_{\rm TM}^2}{2m_{\rm TM}}\,T^{-1}\right)^{\!\!2}}\;
 S_{T,\ GRS}^{1/2}(\omega)
 \label{eq.3g}
\end{equation}

Nominal conditions in the \ltp are the following:

\begin{center}
\begin{tabular}{l}
 $\ell_{\rm TM} = 4.6\times 10^{-2}$ m \\
 ${m_{\rm TM}} = 1.96$ kg \\
 $T = 293$ K \\
 $p = 10^{-5}$ Pa
\end{tabular}
\end{center}

which give
\begin{equation}
 S_{T,\ GRS}^{1/2}(\omega) =
 3.3\times 10^{10}\;S_{a,{\rm thermal\ GRS}}^{1/2}(\omega)\ 
 {\rm K\,Hz}^{-1/2}
 \label{eq.3h}
\end{equation}

This expression gives 0.9\,$\times$\,10$^{-4}$\,K/$\sqrt{\rm Hz}$ in the
worst case that all the thermal acceleration budget, equation~\eref{eq.3a},
is allocated to thermal fluctuations in the {\sl GRS}.

\subsection{Noise effects inside the \textsl{OMS}
\label{sec.11-2}}

The Optical Metrology System of the \ltp is based on a non-polarising,
heterodyne Mach-Zender interferometer~\cite{brax}. The light source is
an infrared Nd:YAG laser of 1.064\,$\mu\/$m wavelength and approximately
0.25~W of power. Optical components are mounted on an optical bench of
highly stable optical properties. Temperature fluctuations basically
affect the optical system through two distinct effects~\cite{defdoc}:

\begin{itemize}
 \item The index of refraction of optical components depends on
	temperature.
 \item Temperature changes cause dilatations (and contractions) of
	optical elements, which in turn cause light's optical path
	to change accordingly.
\end{itemize}

While it is not difficult to characterise how individual components
are influenced by the above effects, to assess the behaviour of the
fully integrated optical metrology is a more complicated task.
Significant progress has been made since the early design proposals
---see reference~\cite{lpfall} and following articles in that issue
of {\sl CQG\/}---, and improved materials and designs are now available.
Altogether, it appears that
%
\begin{equation}
 S_{T,\ {\rm OMS}}^{1/2}\simeq 10^{-4}\,{\rm K}/\sqrt{\rm Hz}
 \label{eq.3i}
\end{equation}
is a sensible requirement which should comfortably guarantee the
performance of the optical bench against temperature fluctuations
in flight ---see~\cite{defdoc}, chapter 12, and~\cite{pete}. Like
before, the noise level~\eref{eq.3i} is estimated to account for about
10\,\% of the total \ltp acceleration noise, equation~\eref{eq.3a}.

\subsection{Total thermal noise budget for the \textsl{LTP}
\label{sec.11-3}}

If we make the hypothesis that noise in the {\sl OMS\/} is uncorrelated
with noise in the {\sl GRS\/} then the respective spectral densities add
up, i.e.,
\begin{equation}
 S_T(\omega) = S_{T,\ GRS}(\omega) +  S_{T,\ OMS}(\omega)
 \label{eq.3j}
\end{equation}

Inserting here equations~\eref{eq.3h} and~\eref{eq.3i}, we find that the
\ltp thermal stability must be
\begin{equation}
 S_T^{1/2}(\omega) \simeq 1.3\times 10^{-4}\,{\rm K}/\sqrt{\rm Hz}
 \label{eq.3k}
\end{equation}
where the usual frequency dependence of equation~\eref{eq.2} has been
dropped, as it is in practice the case that temperature fluctuations
actually drop towards higher frequencies ---due to efficient thermal
insulation of the \ltp.

The number just obtained is very close to that of equation~\eref{eq.3},
and derived under worst hypotheses in each case. We are thus reassured
that equation~\eref{eq.3} is a sensible requirement for the temperature
fluctuations which can be tolerated in~the \ltp. Let us however stress
that we still can count with some margin, as acceleration thermal noise
allocation has been taken as a conservative 10\,\% of the total
acceleration noise ---see paragraph following equation~\eref{eq.3a}.

\section{Temperature measurement and on-ground test requirements
\label{sec.12}}

Temperature stability \emph{requirements} for the \ltp are now established.
We rewrite them again:
\begin{equation}
 S_{T}^{1/2}(\omega)\leq 10^{-4}\,{\rm K}/\sqrt{\rm Hz}\ ,
 \quad 1\,{\rm mHz}\leq \omega/2\pi \leq 30\,{\rm mHz}
 \label{eq.3l}
\end{equation}

Because this is a requirement, satellite and payload will be built such
that thermal conditions in the \ltp meet~\eref{eq.3l}. Even so, to
monitor the magnitude of anyway compliant temperature fluctuations is
a powerful diagnostic tool, yielding valuable information on the system
thermal behaviour, which will be useful for \lisa. In addition, we wish
to be able to diagnose whether the conditions~\eref{eq.3l} actually
prevail at any given time during the mission. For both objectives,
temperature gauges are obviously needed.

We now pose the question: which is the level of noise we can accept in
the temperature measurement system ---which includes sensors, wires and
electronics--- if we are to diagnose temperature variations below the
level~\eref{eq.3l}? Clearly, the answer to that question depends on how
accurately those fluctuations are to be measured. The stability requirement
equation~\eref{eq.3l} is already rather demanding of itself, so we do not
expect thermal fluctuations to be much smaller. With this in mind, to
request the measuring system to be about one order of magnitude less noisy
than the maximum noise level to measure seems a sensible option:
%
%
\begin{equation}
 S_{T,\ {\rm measurement}}^{1/2}(\omega)\leq 10^{-5}\,{\rm K}/\sqrt{\rm Hz}\ ,
 \quad 1\,{\rm mHz}\leq \omega/2\pi \leq 30\,{\rm mHz}
 \label{eq.4}
\end{equation}

This has in fact become a mission top level requirement
---see~\cite{toplev}, section~6.2. There are two groups
of reasons which support it:

\begin{enumerate}
 \item Equation~\eref{eq.3l} defines the \emph{maximum} acceptable
	level of temperature noise in the \ltp. If this is satisfied,
	which of course must, then actual fluctuations will be less
	than that. Requirement~\eref{eq.4} then sets a 10\,\% minimum
	discrimination capability for the measuring device, a standard
	approach which is certainly compatible with better performance.
 \item \lisa is more demanding than \lpf as regards thermal
	stability. Actually, \lisa requires an order of magnitude less
	thermal noise than \ltp~\cite{defdoc}. If we require~\eref{eq.4}
	for \ltp then we are in a position where analysis of thermal
	sources of noise of relevance for \lisa can be identified and
	tagged for improvement, at least in the overlapping frequency
	band of both missions. This prospect is in line with the very
	concept of \lpf as a precursor mission.
\end{enumerate}

\subsection{Absolute versus differential temperature measurements in flight
\label{sec.12-1}}

The above arguments, and specially the first, can perhaps be critisised
in terms of: why not perform \emph{differential} measurements? This might
relax the very demanding requirement in equation~\eref{eq.4}, in the
sense that it would only apply to differential rather than absolute
temperature measurements in flight.

While it is true that certain thermal disturbances depend on temperature
\emph{gradients} across the test masses ---like the radiometer effect and
radiation pressure gradients--- there are others which do not ---mostly
those related to the optics. One could accordingly split up the temperature
gauges into two classes, but this does not seem a particularly sensible
choice, since the best device would obviously be the one to use in all
cases, anyway.

A space mission like \lpf does not generally allow to fix hardware design
inefficiencies once it has been launched. The choice of making applicable
the requirement stated in equation~\eref{eq.4} to \emph{all} temperature
measurements, whether differential or absolute, seems thus not advisable
to relax: some margin is necessary to cope with unforseen sources of error.


\subsection{Environmental conditions for on-ground tests
\label{sec.12-2}}

Before launch, the temperature diagnostics hardware must be tested in
ground. In order to do a meaningful test, a sufficiently stable thermal
environment must be granted for the sensors in the first place. Here,
\emph{sufficiently stable} means that any observed fluctuations in the
readout data should be attributable \emph{solely} to sensor and/or
electronics noise, rather than to a combination of the latter with ambient
temperature fluctuations. This means ambient temperature fluctuations in
the thermometers should be distinctly smaller than the limit set by the
the requirement, equation~(\ref{eq.4}). Again, one order of magnitude
below that target sensitivity seems a good prescription. So we require
\begin{equation}
 S_{T,\ {\rm testbed}}^{1/2}(\omega)\leq 10^{-6}\,{\rm K}/\sqrt{\rm Hz}\ ,
 \quad 1\,{\rm mHz}\leq \omega/2\pi \leq 30\,{\rm mHz}
 \label{eq.5}
\end{equation}

It turns out that 10$^{-6}\,{\rm K}/\sqrt{\rm Hz}$ is a very demanding
temperature stability, orders of magnitude beyond the capabilities of
thermally regulated rooms. Once more, this figure could be widely
relaxed if tests were done \emph{differentially}, i.e., between pairs
of sensors close enough to one another. While such differential
measurements are indeed envisaged, we also want to make sure that
the sensing system is sensitive enough that \emph{absolute} temperature
measurements are compliant with the most exigent requirements. We thus
adopt equation~\eref{eq.5} as the design goal baseline.

\begin{figure}[t]
\centering
\includegraphics[width=6.1cm]{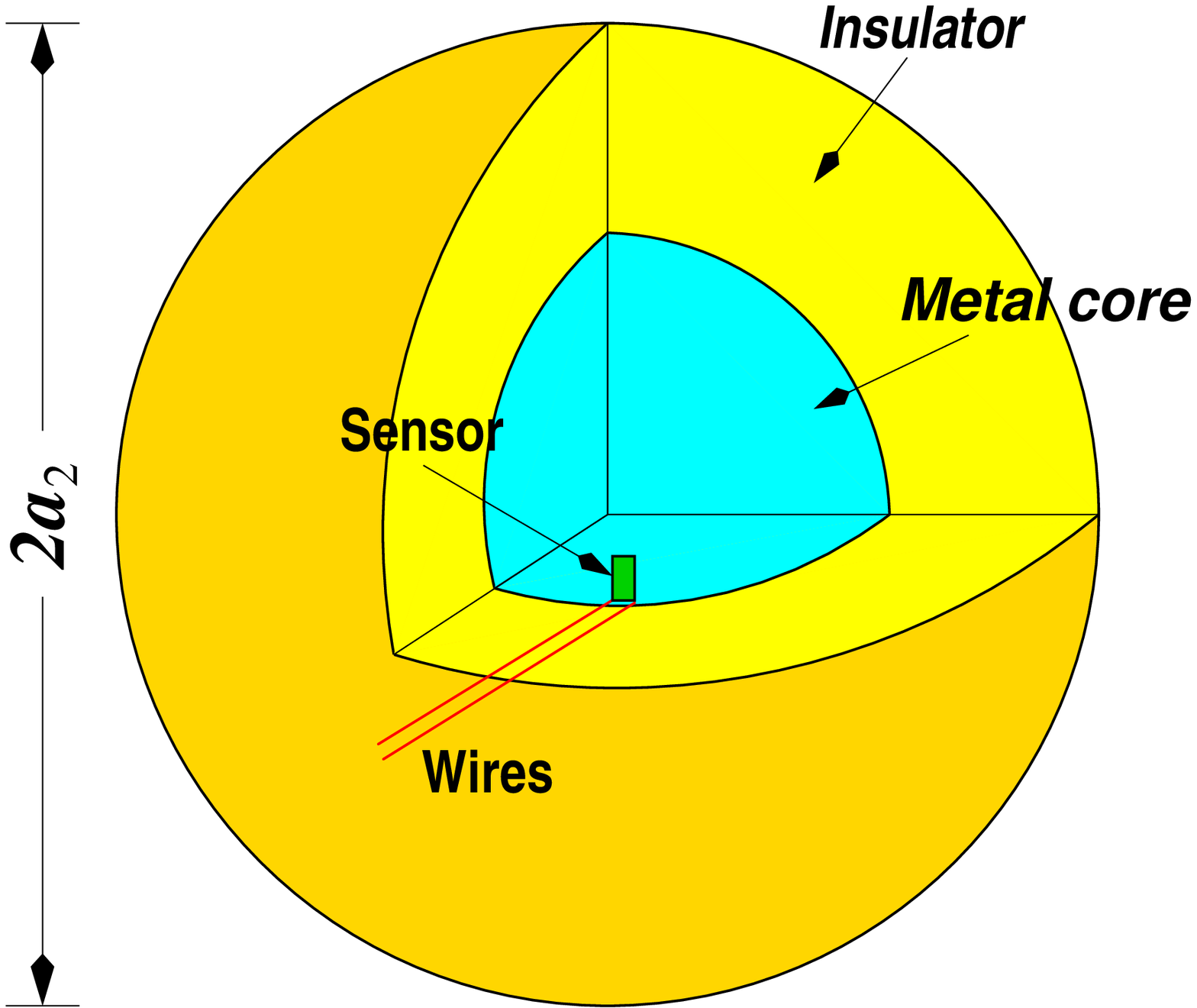}
\qquad
\includegraphics[width=4.9cm]{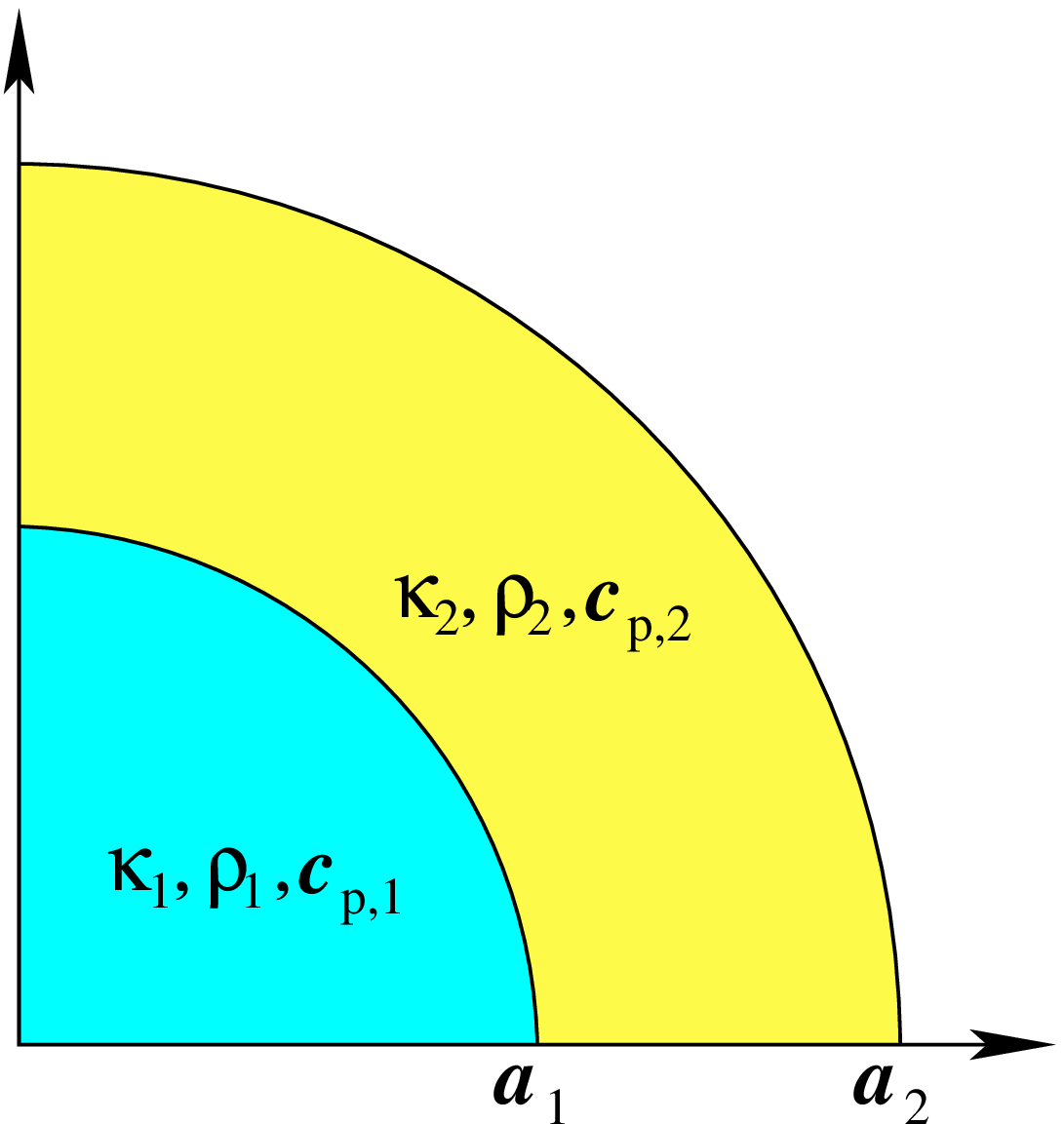}
\caption{Thermal insulator design concept. Left: 3D diagramme, including
sensor placement principle. Right: cut view, with notation convention
dictionary. \label{fig.1}}
\end{figure}

The concept idea of the insulator is displayed in figure~\ref{fig.1}:
an interior metal core of good thermal conductivity is surrounded by
a thick layer of a poorly conductive material. The inner block ensures
thermal stability of the sensors attached to it, while the surrounding
substrate efficiently shields them from external temperature fluctuations
in the laboratory ambient. We propose a spherical shape for the sake
of simplicity of the mathematical analysis, even though this will be
eventually changed to cubic in the actual experimental device for
practical reasons.

\section{Mathematical model of the insulator
\label{sec.2}}

The basic assumption of the mathematical analysis we shall present is
that heat flows from the interior of the insulator to the air outside,
and from the latter to the interior of the insulator, only by thermal
\emph{conduction}. This is a very realistic hypothesis in the context
of the experiment, as radiation mechanisms are certainly negligible
and convection should not play any significant role, either, since
the entire body is solid, and temperature fluctuations will be small
anyway.

Let then $T({\bf x},t)$ be the temperature at time $t\/$ of a point
positioned at vector {\bf x} relative to the centre of the sphere.
Under the conduction-only hypothesis, $T({\bf x},t)$ satisfies Fourier's
partial differential equation~\cite{carslaw}
\begin{equation}
 \rho c_{\rm p}\,\frac{\partial}{\partial t} T({\bf x},t) =
 \nabla\cdot\left[\kappa\nabla T({\bf x},t)\right]
 \label{eq.6}
\end{equation}
where $\rho$, $c_{\rm p}$ and $\kappa$ are the density, specific heat
and thermal conductivity, respectively, of the substrate. We shall
assume these are uniform values within each of the two materials making
up the insulating body, with abrupt changes in the interface. We can
thus represent them as discontinuous functions of the radial coordinate,
as follows:
\begin{equation}
 \rho, c_{\rm p}, \kappa({\bf x}) = \left\{\begin{array}{ll}
 \rho_1, c_{{\rm p}1}, \kappa_1 \quad & {\rm if}\ \ 0\leq r < a_1 \\
 \rho_2, c_{{\rm p}2}, \kappa_2 \quad & {\rm if}\ \ a_1\leq r < a_2
 \end{array}\right.
 \label{eq.7}
\end{equation}
with $r\/$\,$\equiv$\,$|{\bf x}|$. Initial and boundary conditions are
the following:
\begin{equation}
 T({\bf x},t=0) = 0\ ,\quad T(r=a_2,t) = T_0(\theta,\varphi;t)
 \label{eq.8}
\end{equation}
where $\theta\/$ and $\varphi\/$ are spherical angles which define positions
on the sphere's surface. The boundary temperature can be expediently
expressed as a multipole expansion:
\begin{equation}
 T_0(\theta,\varphi;t) = \sum_{lm}\,b_{lm}(t)\,Y_{lm}(\theta,\varphi)
 \label{eq.9}
\end{equation}
where $Y_{lm}(\theta,\varphi)$ are spherical harmonics, and $b_{lm}(t)$
are boundary multipole temperature components.

In practice, the boundary temperature will be \emph{randomly fluctuating},
therefore $b_{lm}(t)$ will be considered \emph{stochastic} functions of
time. We shall also reasonably assume them to be \emph{stationary Gaussian}
noise processes with known spectral densities, $S_{lm}(\omega)$.

As shown in the appendix, the frequency analysis of this problem leads to
a \emph{transfer function} expression of the temperature inside the body:
\begin{equation}
 \tilde T({\bf x},\omega) =
 \sum_{lm}\,H_{lm}({\bf x},\omega)\,\tilde b_{lm}(\omega)
 \label{eq.10}
\end{equation}
where \emph{tildes} (\,$\tilde{}$\,) stand for Fourier transforms, e.g.,
\begin{equation}
 \tilde T({\bf x},\omega)\equiv\int_{-\infty}^\infty\,
 T({\bf x},t)\,e^{-i\omega t}\,dt
 \label{eq.11}
\end{equation}
etc. If we make the further assumption that different multipole temperature
fluctuations at the boundary are \emph{uncorrelated}, i.e.,
\begin{equation}
 \langle\tilde b^*_{l'm'}(\omega)\,\tilde b_{lm}(\omega)\rangle
 = S_{lm}(\omega)\,\delta_{l'l}\,\delta_{m'm}
 \label{eq.12b}
\end{equation}
then the spectral density of fluctuations at any given point inside the
insulating body is given by
\begin{equation}
 S_T({\bf x},\omega) =
 \sum_{lm}\,\left|H_{lm}({\bf x},\omega)\right|^2\,
 S_{lm}(\omega)
 \label{eq.12}
\end{equation}

It is ultimately the spectral density $S_T({\bf x},\omega)$, for
$|{\bf x}|$\,$\leq$\,$a_1$, which has to comply with the requirement
expressed by equation~(\ref{eq.5}). Based on knowledge (by direct
measurement) of ambient laboratory temperature fluctuations,
equation~(\ref{eq.12}) will provide the guidelines, as regards
materials and dimensions, for the actual design of a suitable
insulator jig.

\subsection{Isotropic boundary conditions
\label{sec.2.1}}

Thermal conditions in the laboratory are rather \emph{homogeneous}. This
means that the boundary temperature fluctuations will be in practice
essentially independent of the angles $\theta\/$ and $\varphi$, i.e.,
\begin{equation}
 T_0(\theta,\varphi;t) = B(t)
 \label{eq.13}
\end{equation}
and consequently the generic expansion equation~(\ref{eq.9}) includes
only the \emph{monopole} term. Hence
\begin{equation}
 b_{00}(t) = \sqrt{4\pi}\,B(t)
 \label{eq.14}
\end{equation}

The temperature $T({\bf x},\omega)$ in this case will only depend on
radial depth, $r$, and therefore
\begin{equation}
 \tilde T(r,\omega) = H(r,\omega)\,\tilde B(\omega)
 \label{eq.15}
\end{equation}
with $H(r,\omega)$\,$\equiv$\,$\sqrt{4\pi}\,H_{00}({\bf x},\omega)$.
According to equation~(\ref{eq.a13}) of the Appendix, this is
\begin{equation}
 \hspace*{-0.6 cm}
 H(r,\omega) = \left\{\begin{array}{ll}
 \xi_0(\omega)\,j_0(\gamma_1r)\ , &
 0\leq r \leq a_1 \\[1.2 ex]
 \eta_0(\omega)\,j_0(\gamma_2r) + 
 \zeta_0(\omega)\,y_0(\gamma_2r)\ , &
 a_1\leq r \leq a_2 \end{array}\right.
 \label{eq.16}
\end{equation}

\begin{figure}[t]
\centering
\includegraphics[width=7.8cm]{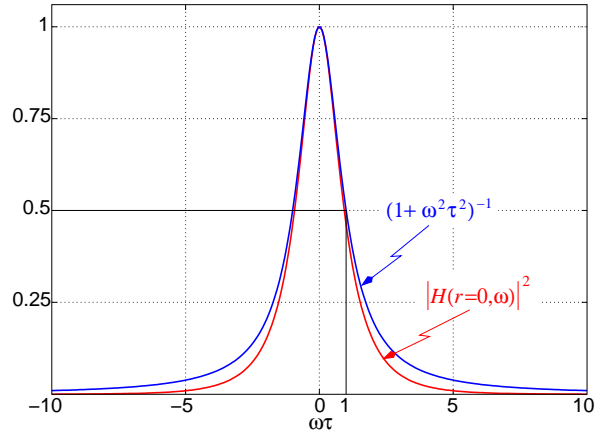}
\caption{Frequency response at the centre ($r\/$\,=\,0) of a spherical
thermal insulator ---as given in equation~(\protect\ref{eq.16})---, along
with the frequency response of a low-pass filter of order one and same
frequency cut-off, $\omega_{\rm cut-off}$\,=\,$1/\tau$. \label{fig.2}}
\end{figure}

This is a low-pass filter transfer function ---even though the
cumbersome frequency dependencies involved in the expressions
above do not make it immediately obvious. The 3\,dB cut-off
angular frequency for this filter defines a time constant~$\tau$
by $1/\tau$\,$\equiv$\,$\omega_{\rm cut-off}$, which correspondingly
is a complicated function of the insulator's physical and geometrical
properties. Figure~\ref{fig.2} shows graphically the situation: the
red curve plots the square modulus of $H(r=0,\omega)$, where the
cut-off $1/\tau$ is shown as the 3\,dB abscissa. For comparison, the
figure also shows (blue curve) a low-pass filter of the first order
with the same frequency cut-off, i.e.,
$|H_{\rm 1st\ order}(\omega)|^2$\,=\,$(1+\omega^2\tau^2)^{-1}$.

The high frequency roll-off of the real filter (in red) is seen to
drop below the first order counterpart (in blue): the latter clearly
has a slow slope at high frequencies,
$|H_{\rm 1st\ order}(\omega)|$\,$\sim$\,$(\omega\tau)^{-1}$, while
the former can be shown to follow a much steeper, exponential curve:
%
%
%
%
%
%
\begin{equation}
 |H(0,\omega)|\sim\omega\tau\,e^{-\sqrt{\omega\tau}}\quad
 {\rm when}\ \ \omega\tau\gg 1
 \label{eq.19}
\end{equation}
%

As already mentioned in section~\ref{sec.1}, to test the temperature
sensors and electronics we need a very strong noise suppression factor
in the \ltp frequency band. Inspection of figure~\ref{fig.2} and
equation~\eref{eq.19} readily shows that high damping factors require
such frequency band to lie in the filter's tails. The thermal insulator
should therefore be designed in such a way that its time constant
$\tau\/$ be sufficiently large to ensure that the \ltp MBW frequencies
are high enough compared to $(2\pi\tau)^{-1}$. The exponential roll-off
in the transfer function shown by~(\ref{eq.19}) makes the filter
actually feasible with moderate dimensions.

\section{Numerical analysis
\label{sec.3}}

In this section we consider the application of the above formalism to
obtain useful numbers for the implementation of a real insulator device
which complies with the needs of the experiment.

First of all, a selection of an \emph{aluminium} core surrounded by a
layer of \emph{polyurethane} was made. Aluminium is a good heat conductor
and is easy to work with in the laboratory; polyurethane is a good
insulator and is also convenient to handle, as it can be foamed to any
desired shape from canned liquid. Other alternatives are certainly
possible, but this appears sufficiently good and we shall therefore
stick to this specific choice. The relevant physical properties of
aluminium and polyurethane are as follows:

\medskip

\begin{center}
\begin{tabular}{lccc}
& Density
& Specific heat
& Thermal conductivity \\
& $\rho$ (kg\,m$^{-3}$)
& $c_{\rm p}$ (J\,kg$^{-1}$\,K$^{-1}$)
& $\kappa$ (W\,m$^{-1}$\,K$^{-1}$) \\[1ex]
\hline \\[-1.3ex]
{\sf Aluminium} & 2700 & 900 & 250 \\
{\sf Polyurethane} & 35 & 1000 & 0.04
\end{tabular}
\end{center}

\medskip

\begin{figure}
\centering
\includegraphics[width=9.7cm]{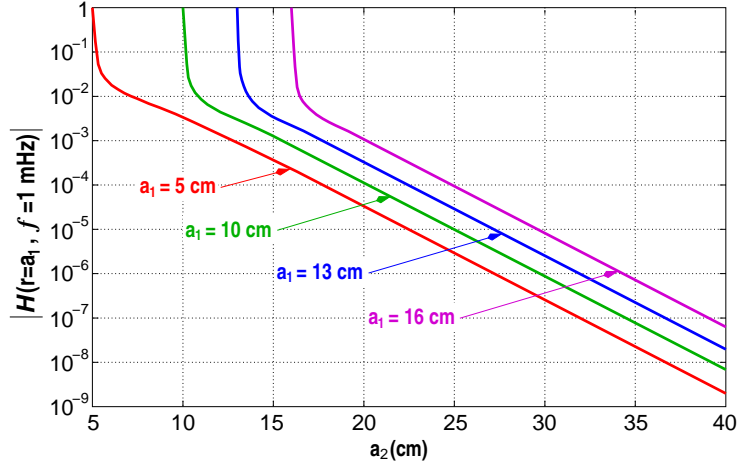}
\caption{Amplitude damping factor on the interface between aluminium
and polyurethane at 1 mHz, for various sizes of the insulating device.
\label{fig.3}}
\end{figure}

Figure \ref{fig.3} plots the \emph{amplitude damping coefficient} of
the insulator block, $|H(r,\omega)|$, at the lower end of the \ltp
frequency band (i.e., 1 mHz) at the interface position, $r\/$\,=\,$a_1$.
Each of the curves corresponds to a fixed value of the latter, and is
represented as a function of the outer radius of the insulator. This
choice is useful because the sensors are implanted for test on the
surface of the aluminium core, and also because at higher frequencies
thermal damping is stronger. The plot clearly shows that the assymptotic
regime of equation~(\ref{eq.19}) is quite early established.

The choice of dimensions for the insulating body must of course ensure
that the minimum requirement, equation~(\ref{eq.5}) is met. For this,
a primary consideration is the size of the ambient temperature
fluctuations in the site where the experiment is made. Dedicated
measurements in our laboratory showed that
\begin{equation}
 S_{T,\ {\rm room}}^{1/2}(\omega)\simeq 10^{-1}\,{\rm K}/\sqrt{\rm Hz}\ ,
 \quad 1\,{\rm mHz}\leq \omega/2\pi \leq 30\,{\rm mHz}
 \label{eq.20}
\end{equation}

We therefore need to implement a device such that
$|H(a_1,\omega)|$\,$\leq$\,10$^{-5}$ throughout the MBW. Suitable
dimensions can then be readily read off figure~\ref{fig.3}, and
various alternatives are possible, as seen. Before making a decission,
however, we need to make an additional estimate of the heat leakage
down the electric wires which connect the temperature sensors with
the electronics, which lies of course outside the insulator. We come
to this next.

\subsection{Heat leakage through connecting wires}

We use a simple model, consisting in assuming the connecting wires
behave as straight thin rods which connect the central aluminium
core with the electronics, placed in the external laboratory ambient.
Because the polyurethane provides a very stable insulation, we can
neglect the lateral flux, hence only a unidimensional heat flow
needs to be considered. For this, the following equation relates
the heat flux to the temperature difference between the two wires'
edges:
\begin{equation}
 \dot{Q}(t) = \kappa_{\rm wire}\,\frac{\pi R_{\rm wire}^2}{\ell_{\rm wire}}\;
 [T(a_2,t)-T(a_1,t)]
 \label{eq.21}
\end{equation}
where $\kappa_{\rm wire}$ is the thermal conductivity of the wire,
$R_{\rm wire}$ its transverse radius, and $\ell_{\rm wire}$ its length
\emph{inside} the polyurethane layer.

On the other hand, the heat flux results in temperature variations in
the metal core, given by
\begin{equation}
 \dot{Q}(t) = \rho_1c_{{\rm p}1}V_1\,
 \frac{\partial T}{\partial t}(a_1,t)
 \label{eq.22}
\end{equation}
where $V_1$\,=\,$4\pi a_1^3/3$ is the volume of the metal core. Equating
the above expressions we find
\begin{equation}
 \kappa_{\rm wire}\,\frac{\pi R_{\rm wire}^2}{\ell_{\rm wire}}\;
 [T(a_2,t)-T(a_1,t)] =
 \rho_1c_{{\rm p}1}V_1\, \frac{\partial T}{\partial t}(a_1,t)
 \label{eq.23}
\end{equation}

For fluctuating temperatures, we can now obtain the relationship between
the spectral density at the aluminium core and the ambient, caused by heat
conduction along the wire:
\begin{equation}
 S_{T,\ {\rm wire}}^{1/2}(a_1,\omega) = |H_{\rm wire}(\omega)|\,
 S_{T,\ {\rm ambient}}^{1/2}(\omega)
 \label{eq.24}
\end{equation}
where
\begin{equation}
 |H_{\rm wire}(\omega)|\simeq\frac{\pi}{\omega}\,
 \frac{\kappa_{\rm wire}\,R_{\rm wire}^2}
 {\rho_1c_{{\rm p}1}V_1\,\ell_{\rm wire}}
 \label{eq.25}
\end{equation}
and where the approximation has been made that the temperature fluctuations
at the inner end of the wire are much smaller than those at the outer
end, due to the presence of the polyurethane layer.

In practice, there will be several sensors for test inside the insulator.
Under the hypothesis made that no lateral heat flux is relevant, the transfer
function for a bundle of $N\/$ of wires is, at most, $N\/$ times that of a
single wire. Thus,
\begin{equation}
 |H_{N{\rm wires}}(\omega)| = \frac{3N}{\omega/2\pi}\,
 \frac{\kappa_{\rm wire}\,R_{\rm wire}^2}
 {8\pi\rho_1c_{{\rm p}1}a_1^3\,\ell_{\rm wire}}
 \label{eq.26}
\end{equation}

Let us consider numerical values in this expression. We use thin
copper wires ($\kappa_{\rm Cu}$\,=\,401\,Wm$^{-1}$K$^{-1}$) of
radius $R_{\rm wire}$\,=\,0.1\,mm, and assume some fiducial
parameters for the size of the aluminium core, $a_1$, the wire
length, $\ell_{\rm wire}$, the number of connecting wires, $N$,
and the frequency, $\omega/2\pi$. The following obtains:
\begin{equation}
 \hspace*{-1.2 cm}
 |H_{N{\rm wires}}(\omega)| = 1.1\times 10^{-5}\,
 \left(\frac{N}{30}\right)\!
 \left(\frac{a_1}{13\ {\rm cm}}\right)^{\!-3}\!
 \left(\frac{\ell_{\rm wire}}{25\ {\rm cm}}\right)^{\!-1}\!
 \left(\frac{\omega/2\pi}{1\ {\rm mHz}}\right)^{\!-1}
 \label{eq.27}
\end{equation}


This result indicates that, for laboratory fluctuations in the level of
equation~(\ref{eq.20}), leakage through wiring causes fluctuations in
the temperature sensors of about 10$^{-6}$\,K/$\sqrt{\rm Hz}$,
equation~\eref{eq.24}, which is compliant with the requirement of
stability, equation~\eref{eq.5}. The most sensitive parameter in
the above expression is the size of the metal core, and this determines
the need to make it somewhat large. The length of the wires has been
taken to be 25~cm, but this does not necessarily mean we need
$a_2$\,=\,38~cm (assuming the radius of the aluminium core is
$a_1$\,=\,13~cm), because the wires can be partly wound inside the
polyurethane layer to further protect the system against leakage.
In fact, wire lengthening is an easy way to improve attenuation.

As regards frequency dependence, compliance is guaranteed in the entire
MBW if it is at its lower end: indeed, not only $|H_{\rm wire}(\omega)|$
decreases as $\omega^{-1}$ ---see equation~\eref{eq.27}\mbox{---,} also
ambient noise fluctuations happen to drop below 10$^{-1}$\,K/$\sqrt{\rm Hz}$
at higher frequencies.

\section{Experimental verification
\label{sec.4}}

Direct experimental verification of the predicted performance of the
insulator would require the sensing system, i.e., sensors plus electronics,
to have a level of noise below 10$^{-6}$\,K/$\sqrt{\rm Hz}$ in the MBW.
The requirement on the latter is however 10$^{-5}$\,K/$\sqrt{\rm Hz}$,
and this is itself to be put to test.

\begin{figure}[b]
\centering
\includegraphics[width=9.7cm,angle=270]{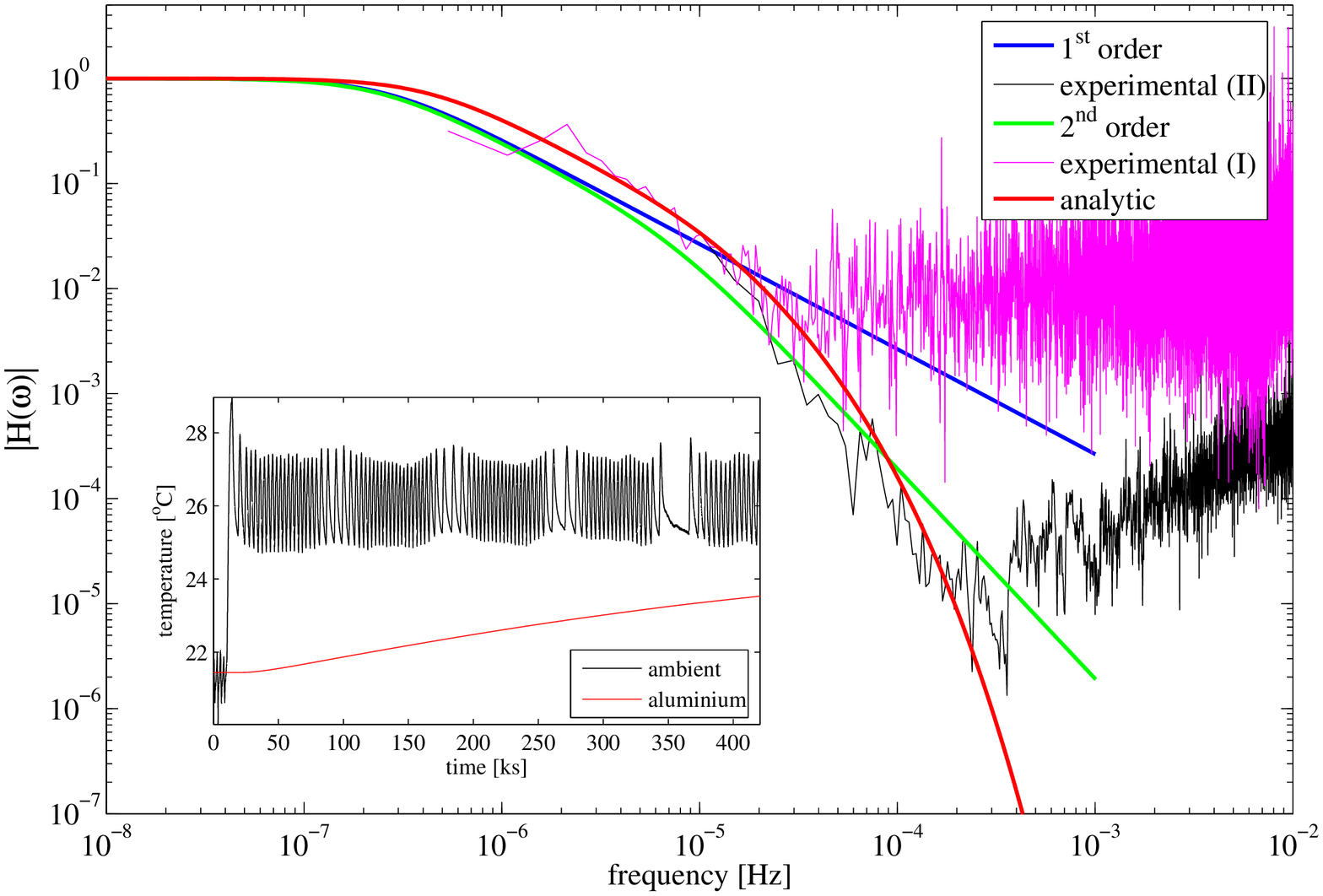}
\caption{Two insulator transfer functions estimated (noisy curves)
and fits to them in the lower frequencies (thicker lines). The inset
contains data for the black curve ---see text for details.
\label{fig.4}}
\end{figure}

We shall nevertheless argue that the thermal environment of the sensors
produced by the insulator is in fact better than required. Evidence of
this comes from rounds of measurements made over periods of several
days to weeks, as we now describe.

Already processed data are displayed in figure~\ref{fig.4}. The noisy
curves are estimates of the insulator transfer function as derived from
two different experimental runs: the magenta curve corresponds to a data
stream about a month long and in largely varying thermal conditions
outside the insulator; the black curve corresponds to a shorter time
series (five days) which began with an abrupt temperature transient,
followed by a periodic signal of a few fractions of a milli-Hertz,
---see figure inset. These curves are relatively clean below
10$^{-4}$\,Hz but they (noisily) tend to approach unity at higher
frequencies, which is an indication that electronic noise is dominant
in that band: indeed, system readouts inside and outside the insulator
tend to equal each other (transfer function nearing unity), while real
temperature fluctuations obviously do not.

The thicker lines are fits to the data in the lower frequency band: the
red is the exact prediction of the model, while the blue is a first order
filter fit, and the green is a second order filter fit. The last two are
provided as examples that the data can also be adjusted by simpler models
in restricted data regions yet indicating that the actual behaviour of the
insulator follows a trend with a steeper slope towards higher frequencies.

Note that the \ltp MBW is above 10$^{-3}$\,Hz, hence is not properly
covered by these data. However the fact that the model predictions are
followed quite well at low frequencies is reassuring, in that we can
expect filter supression factors of 10$^{6}$ and more in the MBW. This
is simply because we of course do not expect the transfer function to
bounce back up again at high frequencies.

The reported facts do not constitute a full quantitative experimental
test of the model, but they do confirm that fluctuation damping is well
below the required level in the MBW.

\section{Conclusions
\label{sec.5}}

Thermal fluctuations in the \ltp must comply with very demanding
requirements, as reflected by equation~\eref{eq.3}. Accordingly, very
delicate sensors and associated electronics must be designed and built
if meaningful temperature measurements are to be performed in flight. The
thermal diagnostics system must then be tested on ground before launch.

However, even the best laboratory conditions fall orders of magnitude
short of the above requirement, so meaningful tests of the temperature
sensing system cannot be tested without suitably screening the sensors
from ambient temperature fluctuations. We have addressed how this can
be accomplished by means of an insulating system consisting of a central
metallic core surrounded by a thick layer of a very poorly conducting
material. The latter provides good thermal insulation, while the central
core, having a large thermal inertia, ensures stability of the sensors'
environment. The choice of materials is flexible, so aluminium and
polyurethane, which are easily available, have been selected.

The appropriate sensors for our needs are temperature sensitive resistors
---more specifically \emph{thermistors}, also known as NTCs. It appears that,
because these sensors need to be wired to external electronics, heat leakage
through such wires is an effect which needs to be quantitatively assessed
as well. We have analysed this effect, and concluded that thermal leakage
depends strongly on the central metallic core size, and requires it to be
somewhat large.

Ambient temperature fluctuations, determined by dedicated \emph{in situ}
measurements, are of the order of 10$^{-1}$\,K/$\sqrt{\rm Hz}$ at 1~mHz
at our laboratory, and decrease at higher frequencies within the MBW.
The required stability conditions at the sensors, attached at the core's
surface, thus need an attenuation factor of 10$^{5}$, or better. Our
analysis determines that a central aluminium core of about 13~cm of
radius, surrounded by a concentric layer of polyurethane 15--20~cm thick,
comfortably provides the needed thermal screening which guarantees a
meaningful test of the sensors' performance.

It could be argued that if differential rather than absolute temperature
measurements were performed in on-ground sensor tests, then a significant
reduction in thermal insulation requirements would ensue. However, the OMS
needs an \emph{absolute} temperature stability of 10$^{-4}$\,K/$\sqrt{\rm Hz}$,
hence an absolute temperature stability measurement on-ground is also needed.
This is why we have chosen the insulator concept described in this paper.

Our results are based on modelling. Direct experimental verification of
the theoretical predictions in the \ltp MBW is however not immediately
obvious, since their successful implementation must result in the
generation of a thermal environment which is more quiet than the
temperature sensing system itself. Nevertheless, analysis of low
frequency real data (i.e., below 1\,mHz), where the insulating
capabilities are less efficient, shows by semi-quantitative extrapolation
that the insulator behaviour in the MBW is in practice much better than
required.

The research presented in this paper specifically applies to the \lpf
mission. But every endeavour regarding \lpf has of course an ultimate
motivation for \lisa. Thermal diagnostics are no exception to the general
rule and, as already pointed out, their design requirements do take into
account that \lisa will be even more exigent as regards thermal stability
and temperature measurements. The results presented in this paper show
that the creation of thermally very stable environments is not a difficult
problem, and that insulator performance can be characterised with great
accuracy by the methods presented here. In fact, the more difficult
problem ---and a major one indeed--- is the design and manufacture of
an electronics which be highly quiet down frequencies of 10$^{-4}$\,Hz
and below, as required for \lisa. We are currently working on these
matters, and will report shortly on progress elsewhere.



\ack

We thank Albert Tom\`as, from {\sl NTE}, for discussions on the insulator
concept. Support for this work came from Project ESP2004-01647 of Plan
Nacional del Espacio of the Spanish Ministry of Education and Science
(MEC). MN acknowledges a grant from Generalitat de Catalunya, and JS
a grant from MEC.

\appendix

\section{Thermal insulator frequency response functions
\label{sec.a1}}

Here we present some mathematical details of the solution to the Fourier
problem, equations~(\ref{eq.6})-(\ref{eq.9}). We first of all Fourier
transform equations~(\ref{eq.6}) and (\ref{eq.9}):
\begin{equation}
 i\omega\,\rho c_{\rm p}\,\tilde T({\bf x},\omega) =
 \nabla\cdot\left[\kappa\nabla\tilde T({\bf x},\omega)\right]
 \label{eq.a1}
\end{equation}

\begin{equation}
 \tilde T_0(\theta,\varphi;\omega) =
 \sum_{l=0}^\infty\sum_{m=-l}^l\,\tilde b_{lm}(\omega)\,Y_{lm}(\theta,\varphi)
 \label{eq.a2}
\end{equation}

Equation (\ref{eq.a1}) can be recast in the form
\begin{equation} 
 \left(\nabla^2 + \gamma_1^2\right)\,\tilde T({\bf x},\omega) = 0\ ,
 \quad 0\leq r\leq a_1
 \label{eq.a3a}
\end{equation}

\begin{equation} 
 \left(\nabla^2 + \gamma_2^2\right)\,\tilde T({\bf x},\omega) = 0\ ,
 \quad a_1\leq r\leq a_2
 \label{eq.a3b}
\end{equation}
where $r\/$\,$\equiv$\,$|{\bf x}|$, and
\begin{equation}
 \gamma_1^2\equiv -i\omega\,\frac{\rho_1 c_{\rm p,1}}{\kappa_1}\ ,\quad
 \gamma_2^2\equiv -i\omega\,\frac{\rho_2 c_{\rm p,2}}{\kappa_2}
 \label{eq.a4}
\end{equation}

To these, matching conditions at the interface\footnote{
The temperature and the \emph{heat flux} should be continuous across
the interface.}
and boundary conditions must be added:
\begin{equation}
 \tilde T(r=a_1-0,\omega) = \tilde T(r=a_1+0,\omega)
 \label{eq.a7a}
\end{equation}

\begin{equation}
 \kappa_1\,\frac{\partial \tilde T}{\partial r}(r=a_1-0,\omega) =
 \kappa_2\,\frac{\partial \tilde T}{\partial r}(r=a_1+0,\omega)
 \label{eq.a7b}
\end{equation}

\begin{equation}
 \tilde T(r=a_2,\omega) = \tilde T_0(\theta,\varphi;\omega)
 \label{eq.a7c}
\end{equation}

Equations (\ref{eq.a3a}) and (\ref{eq.a3b}) are of the Helmholtz kind. Their
solutions are thus respectively given by
\begin{equation}
 \hspace*{-2.25 cm}
 \tilde T({\bf x},\omega) = \left\{\begin{array}{ll}
 \displaystyle
 \sum_{lm}\,A_{lm}(\omega)\,j_l(\gamma_1r)\,Y_{lm}(\theta,\varphi)\ , &
 0\leq r \leq a_1 \\[1.7 em]
 \displaystyle
 \sum_{lm}\,\left[C_{lm}(\omega)\,j_l(\gamma_2r) + 
                  D_{lm}(\omega)\,y_l(\gamma_2r)\,\right]\,
                  Y_{lm}(\theta,\varphi)\ , &
 a_1\leq r \leq a_2 \end{array}\right.
 \label{eq.a5}
\end{equation}
where $j_l\/$ and $y_l\/$ are spherical Bessel functions \cite{as72},
\begin{equation}
 \hspace*{-0.8 cm}
 j_l(z) = z^l\,\left(-\frac{1}{z}\,\frac{d}{dz}\right)^{\!\!l}\,
 \frac{\sin z}{z}\ ,\quad
 y_l(z) = -z^l\,\left(-\frac{1}{z}\,\frac{d}{dz}\right)^{\!\!l}\,
 \frac{\cos z}{z}
 \label{eq.a6}
\end{equation}
and the coefficients $A_{lm}(\omega)$, $C_{lm}(\omega)$ and $D_{lm}(\omega)$
are to be determined by equations~(\ref{eq.a7a})--(\ref{eq.a7c}). These can
be expanded as follows, respectively:
\begin{eqnarray}
 \sum_{lm}\,A_{lm}(\omega)\,j_l(\gamma_1a_1)\,Y_{lm}(\theta,\varphi)\ =
 & & \nonumber \\
 \ \ =\ \sum_{lm}\,\left[C_{lm}(\omega)\,j_l(\gamma_2a_1) +
                  D_{lm}(\omega)\,y_l(\gamma_2a_1)\,\right]\,
                  Y_{lm}(\theta,\varphi) & &
 \label{eq.a8a}
\end{eqnarray}

\begin{eqnarray}
 \kappa_1\gamma_1\,
 \sum_{lm}\,A_{lm}(\omega)\,j'_l(\gamma_1a_1)\,Y_{lm}(\theta,\varphi)\ =
 & & \nonumber \\
 \ \ =\ \kappa_2\gamma_2\,
 \sum_{lm}\,\left[C_{lm}(\omega)\,j'_l(\gamma_2a_1) +
                  D_{lm}(\omega)\,y'_l(\gamma_2a_1)\,\right]\,
                  Y_{lm}(\theta,\varphi)
 \label{eq.a8b}
\end{eqnarray}

\begin{eqnarray}
 \sum_{lm}\,\left[C_{lm}(\omega)\,j_l(\gamma_2a_2) + 
                  D_{lm}(\omega)\,y_l(\gamma_2a_2)\,\right]\,
                  Y_{lm}(\theta,\varphi)\ =
 & & \nonumber \\
 \ \ =\ \sum_{lm}\,\tilde b_{lm}(\omega)\,Y_{lm}(\theta,\varphi)
 \label{eq.a8c}
\end{eqnarray}

Because of the completeness property of the spherical harmonics, the
above equations completely determine the coefficients $A_{lm}(\omega)$,
$C_{lm}(\omega)$ and $D_{lm}(\omega)$. The result is
\begin{equation}
 \hspace*{-2 cm}
 A_{lm}(\omega) = \xi_l(\omega)\,\tilde b_{lm}(\omega)\ ,\ \ 
 C_{lm}(\omega) = \eta_l(\omega)\,\tilde b_{lm}(\omega)\ ,\ \ 
 D_{lm}(\omega) = \zeta_l(\omega)\,\tilde b_{lm}(\omega)
 \label{eq.a9}
\end{equation}
with
\begin{equation}
 \hspace*{-1 cm}
 \xi_l(\omega) = \frac{1}{\Delta_l(\omega)}\,\left[
 \kappa_2\gamma_2\,j_l(\gamma_2a_1)\,y'_l(\gamma_2a_1) -
 \kappa_2\gamma_2\,j'_l(\gamma_2a_1)\,y_l(\gamma_2a_1)\right]
 \label{eq.a9a}
\end{equation}

\begin{equation}
 \hspace*{-1 cm}
 \eta_l(\omega) = \frac{1}{\Delta_l(\omega)}\,\left[
 \kappa_2\gamma_2\,j_l(\gamma_1a_1)\,y'_l(\gamma_2a_1) -
 \kappa_1\gamma_1\,j'_l(\gamma_1a_1)\,y_l(\gamma_2a_1)\right]
 \label{eq.a9b}
\end{equation}

\begin{equation}
 \hspace*{-1. cm}
 \zeta_l(\omega) = \frac{1}{\Delta_l(\omega)}\,\left[
 \kappa_1\gamma_1\,j_l(\gamma_2a_1)\,j'_l(\gamma_1a_1) -
 \kappa_2\gamma_2\,j'_l(\gamma_2a_1)\,j_l(\gamma_1a_1)\right]
 \label{eq.a9c}
\end{equation}
and
\begin{eqnarray}
 \hspace*{-1 cm}
 \Delta_l(\omega) & = &
 \ \kappa_1\gamma_1\,j'_l(\gamma_1a_1)\,\left[
 j_l(\gamma_2a_1)\,y_l(\gamma_2a_2) -
 j_l(\gamma_2a_2)\,y_l(\gamma_2a_1)\right]\ + \nonumber \\
 & + &
 \ \kappa_2\gamma_2\,j_l(\gamma_1a_1)\,\left[
 j_l(\gamma_2a_2)\,y'_l(\gamma_2a_1) -
 j'_l(\gamma_2a_1)\,y_l(\gamma_2a_2)\right]
 \label{eq.a10}
\end{eqnarray}

When the above results, equations~(\ref{eq.a9a}) through (\ref{eq.a10}),
are inserted back into equation~(\ref{eq.a5}) the result stated in
equation~(\ref{eq.10}) in the main text obtains, i.e.,
\begin{equation}
 \tilde T({\bf x},\omega) =
 \sum_{lm}\,H_{lm}({\bf x},\omega)\,\tilde b_{lm}(\omega)
 \label{eq.a11}
\end{equation}
where
\begin{equation}
 \hspace*{-1.8 cm}
 H_{lm}({\bf x},\omega) = \left\{\begin{array}{ll}
 \xi_l(\omega)\,j_l(\gamma_1r)\,Y_{lm}(\theta,\varphi)\ , &
 0\leq r \leq a_1 \\[1.2 ex]
 \left[\eta_l(\omega)\,j_l(\gamma_2r) + 
                  \zeta_l(\omega)\,y_l(\gamma_2r)\,\right]\,
                  Y_{lm}(\theta,\varphi)\ , &
 a_1\leq r \leq a_2 \end{array}\right.
 \label{eq.a12}
\end{equation}

For monopole only boundary conditions, equation~(\ref{eq.15}), the
transfer function is
\begin{equation}
 \hspace*{-0.6 cm}
 H(r,\omega) = \left\{\begin{array}{ll}
 \xi_0(\omega)\,j_0(\gamma_1r)\ , &
 0\leq r \leq a_1 \\[1.2 ex]
 \eta_0(\omega)\,j_0(\gamma_2r) + 
                  \zeta_0(\omega)\,y_0(\gamma_2r)\ , &
 a_1\leq r \leq a_2 \end{array}\right.
 \label{eq.a13}
\end{equation}


\end{document}